\documentclass[prb,aps,twocolumn]{revtex4}
\usepackage{epsf}
\usepackage{epsfig}
\usepackage{graphics}

\begin{document}

\title{Structural Ordering and Antisite Defect Formation 
in Double Perovskites}

\author{Prabuddha Sanyal, Sabyasachi Tarat and Pinaki Majumdar}

\affiliation{Harish-Chandra  Research Institute,
 Chhatnag Road, Jhusi, Allahabad 211 019, India}

\date{9 April  2008}

\begin{abstract}
We formulate an effective model for B-B' site ordering in double perovskite 
materials A$_2$BB'O$_6$. Even within the simple framework of lattice-gas type 
models, we 
are able to address several experimentally observed issues including 
nonmonotonic dependence of the degree of order on annealing temperature, and 
the rapid decrease of order upon overdoping with either B or B' species. We also 
study ordering in the `ternary' compounds A$_2$BB'$_{1-y}$B''$_y$O$_6$. 
Although our emphasis is on the double perovskites, our results are easily 
generalizable to a wide variety of binary and ternary alloys.  
\end{abstract}

\maketitle

\section{Introduction}

Double perovskite (DP) materials of the form A$_2$BB'O$_6$
are being actively studied \cite{dp-rev} on account of their 
unusual electronic and magnetic properties. 
In particular, some double perovskites, {\it e.g}, Sr$_2$FeMoO$_6$,
show half-metallic behaviour,  high ferromagnetic $T_c$, 
and large low field magnetoresistance.
Remarkably, there are also double perovskites that are
antiferromagnetic and insulating. 
The metallic and magnetic character depends primarily on the
choice of B and B', but is also affected by the
{\it local ordering} of these ions.

While it is desirable to understand the 
interplay of structural, electronic and magnetic variables
in these materials within an unified framework, it 
is technically daunting to `anneal' all these degrees
of freedom simultaneously. 
We argue that it is reasonable to abstract the problem of 
structural order, solve it, and then 
set up the electronic-magnetic problem on the 
appropriate structural motif.  
Our paper is organised as follows. In the next section we
present the overall model for the DP's, involving the
structural, magnetic and electronic degrees of freedom,
and indicate how an effective structural model can be
formally extracted. The section after, discusses a phenomenological
model for structural order.
We then discuss our method and computational variables. 
This is followed by the results,
first on `ordering' for materials of the form
A$_2$BB'O$_6$, then for the `doped' cases, 
A$_2$B$_{1+x}$B'$_{1-x}$O$_6$, and finally for the
ternary systems, 
A$_2$BB'$_{1-y}$B''$_y$O$_6$. We conclude with a
summary of our results.

\section{Structural order and magnetism}

The magnetism in the DP's is intimately related to electron
delocalisation, which in turn depends on the spatial pattern
of B, B' ions~\cite{KimCheong,Martinez,GarciaLanda}. Let us write down the Hamiltonian in terms
of the electronic and magnetic degrees of freedom to
illustrate the crucial role of spatial B-B' order.
Let B be the `magnetic' ion and B' the non magnetic one.
We define a binary variable  $\eta_i$, such that
$\eta_i =1$ when a site has a B ion, and $\eta_i =0$ when
a site has a B' ion. The $\eta$'s will encode the
atomic positions. In terms of these, the
model for the DP's~is:

\begin{eqnarray}
H && = \epsilon_{B}\sum_i \eta_i f_{i\sigma}^{\dagger}f_{i\sigma}+
\epsilon_{B'}\sum_i (1 - \eta_i) m_{i\sigma}^{\dagger}m_{i\sigma} \cr
&&~ -t\sum_{<ij>\sigma} \eta_i (1 - \eta_j) 
f_{i\sigma}^{\dagger}m_{j\sigma} \cr
&&~+ J\sum_i \eta_i {\bf S}_{i} \cdot
f_{i\alpha}^{\dagger}\vec{\sigma}_{\alpha\beta}f_{i\beta} 
+ J_{AF} \sum_{\langle ij \rangle} \eta_i \eta_j {\bf S}_i.{\bf S}_j \cr
&&~+ V_{at}\{ \eta \}
\end{eqnarray}

The $f$ operators  refer to the magnetic B sites and 
the $m$ to the non magnetic B'. $\epsilon_{B}$ and 
$\epsilon_{B'}$ are level energies, respectively, at the
B and B' sites,  $\Delta =\epsilon_{B} - \epsilon_{B'}$
is the `charge transfer' energy, and $t$ is the hopping amplitude
between nearest neighbour B and B' ions. We have ignored
orbital degeneracy in the present model.
${\bf S}_i$ are the
moments on the B site, $J$ is the Hunds
coupling on those sites, and  $J_{AF}$ is the antiferromagnetic (AF)
superexchange coupling between  B moments when two B ions
neighbour each other. The $V_{at}\{ \eta \}$ represent
atomic interactions between the B, B' ions.

In the simplest case of equal proportions of  B and B' ions,
and their perfect alternating arrangement, each B is coordinated 
by B' only and vice versa. The AF coupling
does not come into play, and electron delocalisation on
the B-B' network generally prefers a ferromagnetic spin 
configuration.
This state is also highly conducting. However, if the atomic
order is imperfect and there are B ions neighbouring each other,
two B moments get locked into an antiparallel configuration.
This leads to  a reduction in
the overall ferromagnetic moment, and these `antisite' regions
also hinder electron transport. The magnetism and
transport is obviously strongly dependent on the structural
motif. What decides the atomic B-B' arrangement?
Let us look at the formal answer first.

For large $S$ spins, the DP model 
refers to electrons coupled to classical magnetic
moments and moving in a background defined by $\{ \eta \}$.
If the atomic ordering problem is to be isolated from this,  
one should `trace out' the electronic and magnetic
variables. The effective potential $V_{eff} \{ \eta \}$
controlling atomic order 
is
$$
V_{eff} \{ \eta \}
= -(1/\beta) log\int {\cal D} {\bf S}_i 
Tr_{\{f,m\}} e^{-\beta H}
$$
Computing $V_{eff}$ and `updating' atomic positions
accordingly is a computationally demanding task, the
Monte Carlo (MC) equivalent of a Car-Parinello simulation.
There is limited information about $V_{at}$, and the
`trace' is technically difficult, so 
we construct a simple
$V_{eff} \{ \eta \}$
that is consistent with the phenomenology, rather than
attempt an elaborate `first principles' calculation.
We motivate this in the next section.

\section{Model for atomic ordering}

The ideal DP with the general formula A$_2$BB'O$_6$ 
has ordering of B and B' atoms at the center of
alternate O$_6$ octahedra. Focusing only on the
B and B' ions, the ideal structure is simply
an alternation
of B and B' ions along the ${\hat x}$, ${\hat y}$ and
${\hat z}$ axis.
However, in imperfectly annealed systems,
there can occur antisite defect regions where this 
ordering is reversed, and two B atoms or two 
B' atoms occur adjacent to each other. 
The ordering of the B-B' is in general neither
`perfect' nor random, it is the result of an
annealing process. While samples with
high degree of order have been grown~\cite{Kobayashi}, indicating
that the atomic order in the structural ground state
should be perfect, recent 
experiments reveal an interesting trend in
the degree of order as a function of annealing 
temperature.

In the experiments  by  Sarma 
{\it et. al.}~\cite{ord-expt-prl} on Sr$_2$FeMoO$_6$ 
it was observed that there was a non-monotonic 
dependence of the degree of Fe-Mo
ordering upon the annealing temperature, $T_{ann}$,
as shown in 
Fig.1.
The different samples were taken from the same `parent'
material (synthesised at high temperature and quenched to a
low temperature), heated to a temperature $T_{ann}$, and
annealed there for a duration $\tau_{ann}$, say. 

If there is indeed a B-B' ordering tendency in the DP's,
the extent of order {\it at  equilibrium} would be
highest at low $T_{ann}$, progressively falling off
at higher $T_{ann}$ where the order is expected to be
small.
The downturn at low annealing temperature, it seems
reasonable, 
is due to insufficient equilibriation. 
Our model below and results are based 
on the assumption that: (i)~there is an intrinsic
B-B' ordering tendency in the DP's, (ii)~given long
equilibriation time, the DP's would indeed show a
high degree of order at low $T$, but (iii)~under
typical synthesis conditions/annealing protocol
the system only manages to generate correlated
configurations with short range order. The 
annealing temperature and annealing time are
therefore key to quantifying the structural order.

Since the structural (dis)order seems to be
`frozen' at temperature, $T \sim 1000$K,  much
above the temperature for magnetic order ($\sim 400$K), 
the qualitative issues in atomic ordering can be
understood by ignoring the electronic-magnetic variables
in an effective model, below.

\begin{figure}[h]
\vspace{.2cm}
\centerline{
\includegraphics[width=7.2cm,height=5.0cm,clip=true]{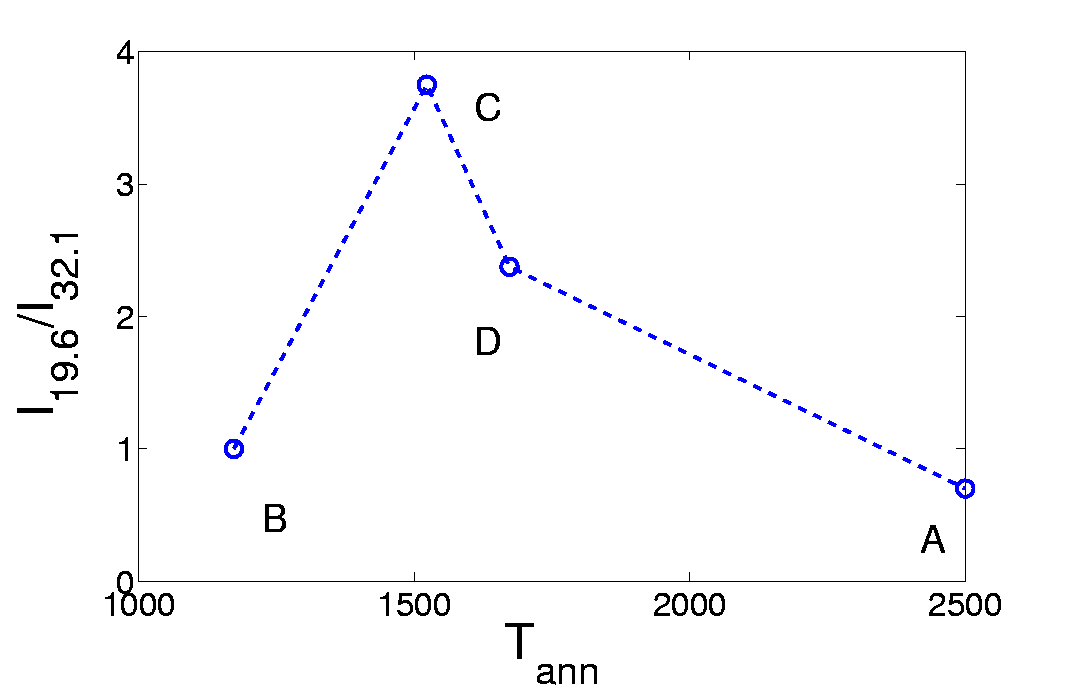}}
\vspace{.2cm}
\caption{Experimental data for ordering vs annealing
temperature, obtained by Sarma {\it et. al}}
\end{figure}

In the absence of detailed microscopic knowledge about 
$V_{eff} \{ \eta \}$, we used a binary lattice gas
model that has the same ordering tendency as the real
materials, {\it viz}, B-B' alternation, or equivalently
a `checkerboard' pattern.
In terms of the variables $\eta_i$, the simplest such 
model is:
\begin{equation}
V_{eff} \{ \eta \} = V \sum_{\langle ij\rangle} \eta_i (1 - \eta_j)
\end{equation}
with  $V > 0$ being a measure of the ordering tendency. The
ground state in this model would correspond to $\eta =1,0,1,0,..$
along each axis, {\it i.e}, B, B', B, B'..
Notice that this approach tries to incorporate the
effect of complex interactions 
between the A, B, B' and O ions (as also the electrons) into 
a single parameter (See Appendix).

This binary model is equivalent to a nearest neighbour
Ising model, so the {\it equilibrium physics} is very well
understood~\cite{staufferbook}. We, however, want to explore the consequences of
different annealing protocols on this model, to examine
the consequences of imperfect annealing. 
The qualitative ordering effects in the lattice gas
model are similar in 2D and 3D, so 
we work with a 2D structure
since it allows ease of visualisation.

\section{Method}
The atomic order that emerges from $V_{eff}$ can be studied
with mean field theory if the system is in equilibrium.
However, we want to explore non equilibrium effects due
to poor annealing, so we use a Monte Carlo technique to
anneal the $\eta$ variables. Most of our studies involve a
protocol where we start with a completely random B-B'
configuration (as if quenched from very high $T$), and
then anneal it at a temperature $T_{ann}$, for a MC run
of duration 
$\tau_{ann}$ Monte Carlo steps (MCS). 
 Since the number of B and B' atoms is fixed
we update our configurations by (i)~moving to some site
${\bf R}_i$, (ii)~attempting an {\it exchange} of the
atom at ${\bf R}_i$, with another randomly picked within
a box of size $L_C^2$ centred on ${\bf R}_i$, and (iii)~accepting
or rejecting the move based on the Metropolis algorithm.
We have used system size $L=20,~ 40,$ and $80$, update cluster
with $L_C=4,$ and $10$, 
$T_{ann}/V$ ranging from $0.01-0.3$, and $\tau_{ann}
\sim 200-30000$ MCS.
We have studied the structure factor, and also detailed
spatial configurations, to quantify the extent of order.

\section{Ordering with equal proportions of B, B'}

This section discusses the order in systems where
the number of B and B' ions is equal, and perfect order
is in principle possible.
In an attempt to mimic the experimental annealing protocol, we
started with random initial configurations at some temperature
$T_{ann}$ and annealed for some time $\tau_{ann}$. 
The structure
factor $S({\bf q})$ 
at the ordering wavevector, $\{\pi, \pi\}$, is averaged
over 40 such initial configurations. 
The extent of order, quantified by the 
peak in $S({\bf q})$, is shown in Fig.2(a)
for a lattice size $20 \times 20$, and in Fig.2(b), for
a lattice size $40 \times 40$. 

There are two noteworthy features in the data in Fig.2. (i)~The
non monotonic behaviour with $T_{ann}$ that we anticipated indeed 
shows up in the structure factor peak, and (ii)~there is a 
strong size dependence of the peak in  $S ( \pi, \pi)$, varying
almost by a factor of 4 between $L=20$ and $L=40$!

\begin{figure}[b]
\centerline{
\includegraphics[width=7.0cm,height=8.0cm,angle=0,clip=true]{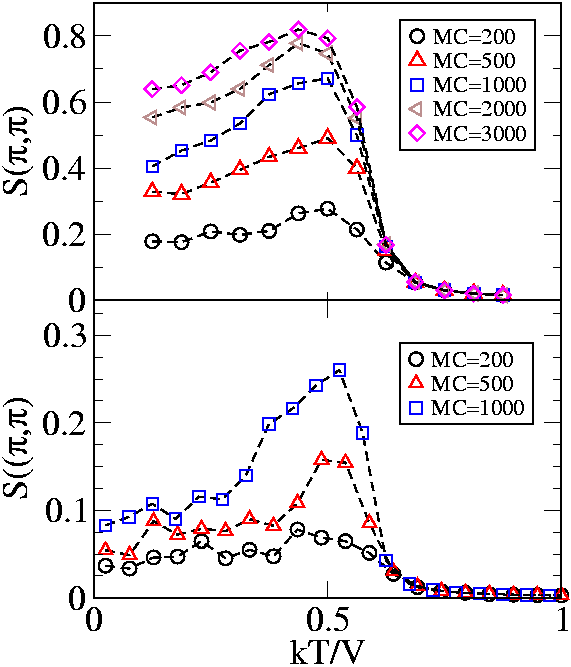}}
\vspace{.2cm}
\caption{$\pi-\pi$ structure factor vs $T_{ann}$
on a $20 \times 20$ lattice,
for different Monte Carlo runlength (annealing time).
$20 \times 20$ lattice (top) and $40 \times 40$ lattice (bottom).}
\end{figure}

\begin{figure}[t]
\centerline{
\includegraphics[width=5.0cm,height=5.0cm,clip=true]{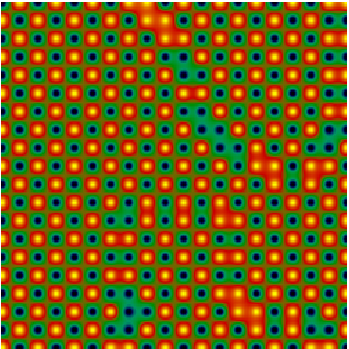}}
\caption{Non-equilibrium, $T_{ann} = 0.5$ configuration for
$\tau_{ann} =100$.}
\end{figure}

The downturn in  $S ( \pi, \pi)  $ at low $T_{ann}$ is due
to the inability of the system to achieve equilibrium, at
short annealing time, when one starts with a random initial
configuration. With increasing $\tau_{ann}$ there is of 
course an increase in the extent of order (for given $T_{ann}$
and $L$) and $S(\pi, \pi)$ reaches $\sim 0.6$ of 
the equilibrium value for
$L=20$, for MCS=1000. At $L=40$, however, this is suppressed to $\sim 0.1$ of 
the equilibrium value, for the same MCS. This origin of this strong size
dependence  
becomes apparent when we examine a typical
configuration 
at low $T_{ann}$,
generated by a short annealing run, $\tau_{ann}
= 100$ MCS, in Fig.3

The ordering is obviously 
imperfect, as $S ( \pi, \pi )$ suggested, more
interestingly, the system actually consists of a few 
large ordered clusters with `phase slip' between them.
While locally these domains are well ordered, the bulk
$S ( \pi, \pi )$ arises from the interfering
contribution of large domains, and this `cancellation'
depends  strongly on the system size $L$.
In the smaller systems, $L \sim 20$,
$S ( \pi, \pi )$ is decided by
the larger domain, and as domains proliferate with increasing
$L$, there is an increasingly better cancellation between
the out of phase domains, and $S(  \pi, \pi) $ falls.

Since the antisite regions are at the interface of two
ordered (but phase slipped) clusters, in what follows we
show
the domains and domain walls, rather than the detailed
atomic configuration.

\begin{figure}[t]
\vspace{.2cm}
\centerline{
\includegraphics[width=7.5cm,height=10cm,angle=0,clip=true]{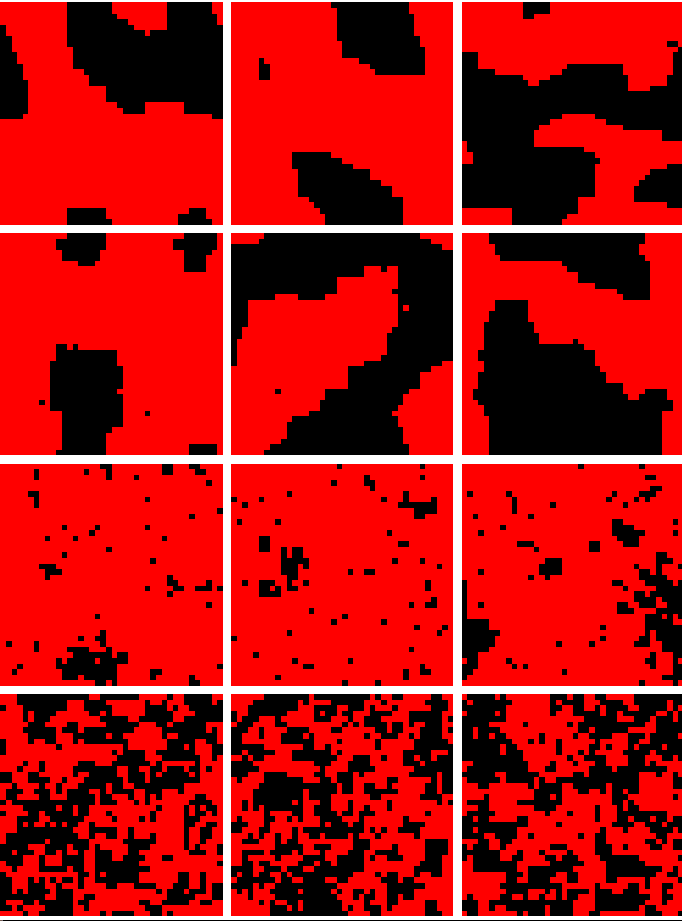}}
\vspace{.2cm}
\caption{Domain pattern for different $T_{ann}$ and
equilibriation time. Left column, $\tau_{MC} = 1000$, center
$\tau_{MC} = 500$, right $\tau_{MC} = 200$. Rows, top to bottom,
$T =0.1,~1.0,~2.0,~3.0$.
The antisite defects are at the domain boundary.}
\end{figure}

In Fig.4 one corner site in each panel is set as reference and 
the others `coloured' in terms of their phase relation to it.
For concreteness assume the atom at bottom left 
corner is B. Let this site be ${\bf R}_0 =\{0,0\}$, and index
all sites in terms of integers, $\{R_{ix}, R_{iy}\}$. Then the
`correct' order would require that the atom at ${\bf R}_i$
be B if  $R_{ix} + R_{iy}$ is even, and B' if  
$R_{ix} + R_{iy}$ is odd. 
Formally, we just plot $f(\eta_i) =
\eta_i e^{i \pi (R_{ix} + R_{iy})}$, 
with a light colour for a site that is correctly ordered with respect
to the origin, and dark  if it is not.

The result of this, on a $40 \times 40$ lattice, is shown in
Fig.4. The columns correspond to different $\tau_{ann} =1000,~500,~200$ MCS, 
left to right, and the rows to $T_{ann}=0.1,~1.0,~2.0,~3.0$, top to
bottom.
At larger annealing and intermediate $T_{ann}$ we see 
a large single domain dominating the configuration, while
at low or high $T_{ann}$ and short $\tau_{ann}$ the structure
is more fragmented.
In the electron problem, the domains themselves are likely to be
`homogeneous' ferromagnetic regions, while the antisite
regions, with neighbouring B-B atoms might be antiferromagnetic.

\begin{figure}[b]
\vspace{.2cm}
\centerline{
\includegraphics[width=6.0cm,height=5.0cm,angle=0,clip=true]{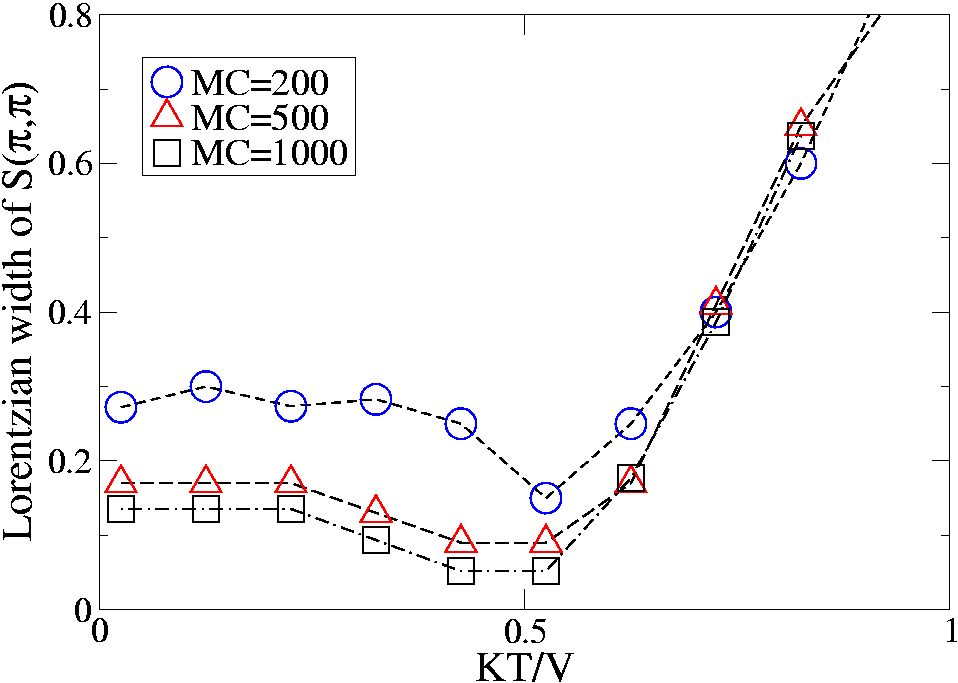}}
\vspace{.2cm}
\caption{The Lorentzian width in the structure factor peak, to estimate
the domain size/correlation length.}
\end{figure}

In an attempt to quantify the size of the clusters as a function
of $T_{ann}$ and $\tau_{ann}$ we studied the full structure factor
$S({\bf q})$, averaged over $\sim 50$ configurations, and 
fitted the ordering peak  to a lorentzian of the form
$$
S({\bf q} \sim \{\pi, \pi\}) \approx
A (\Gamma/\pi)/( (q_x - \pi)^2 + (q_y - \pi)^2 + \Gamma^2)
$$
$A$ is an overall amplitude factor and 
$\Gamma(T_{ann}, \tau_{ann})$ is a measure of the (inverse)
width of the cluster size. This is shown in Fig.5. 

We also tried a protocol where a system is heated to a
temperature $T_f$ but gradually, through a sequence
$T_1,~T_2,..$, retaining the memory of configurations
at the earlier temperature. The result of this slow
annealing is similar to the earlier protocol,
{\it i.e}, non-monotonic for short $\tau_{ann}$,
and tends towards the monotonic equilibrium response
with increasing $\tau_{ann}$. For $\tau_{ann} > 10^4$,
we obtain essentially the equilibrium profile except
at very low $T$. The results are in Fig.6.

\begin{figure}[t]
\vspace{.2cm}
\centerline{
\includegraphics[width=6.0cm,height=4.5cm,angle=0,clip=true]{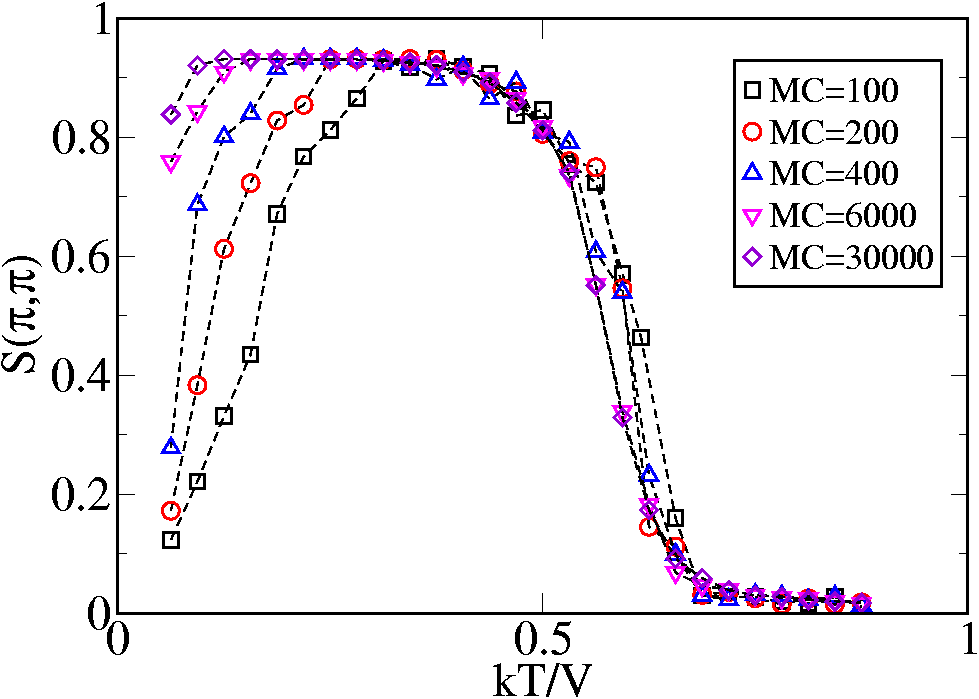}}
\vspace{.2cm}
\caption{$\pi-\pi$ structure factor vs $T$ for heating run with memory for
different Monte Carlo steps (annealing time)}
\end{figure}


\section{Ordering with unequal proportions of B, B'}

When the B-B' proportion is not 1:1 it is obviously
not possible for every B to be coordinated by B' and
{\it vice-versa}. `Antisite' regions, rich in B or B',
would exist even at equilibrium. Our attempt, in this
section is (i)~to capture the loss of B-B' order with
increasing concentration of B', say, and also (ii)~to
study the effect of restricted annealing on the system.

Our reference for (i)~above are the results of
D. Topwal, {\it et. al.}~\cite{topwal}, who  had 
prepared the series
of samples Sr$_{2}$Fe$_{1+x}$Mo$_{1-x}$O$_{6}$, with  
proportion of Fe  in excess of Mo.
Such samples would have 
antisite defects even when the sample is well
equilibriated.  The XRD pattern for 
these set of samples is given
in Fig.7. They observed that the peak at 
$19.6^{\circ}$, related 
to checkerboard
ordering, decreases uniformly for both positive as well as 
negative $x$. In particular, the
peak is observed to be almost absent in the $x=0.5$ sample.  
 
\begin{figure}[h]
\centerline{
\includegraphics[width=6.0cm,height=5.0cm,clip=true]{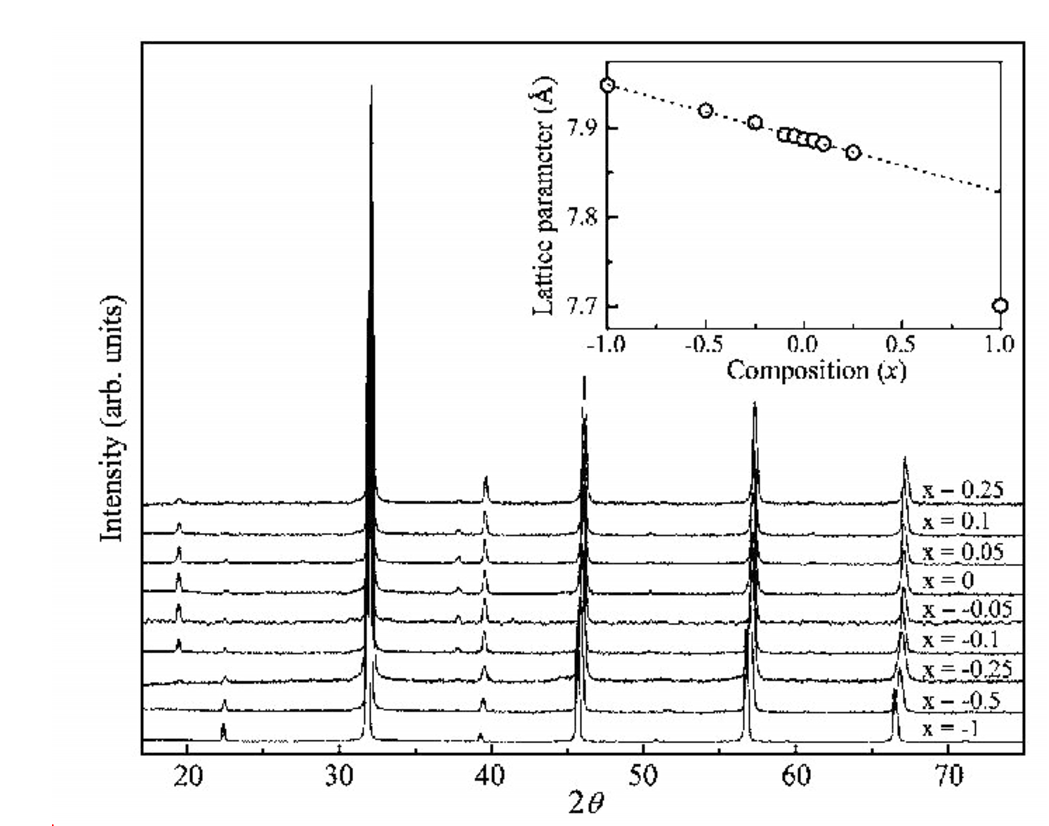}}
\caption{
XRD data for different Fe-Mo proportions,
obtained by Topwal {\it et. al}~\cite{topwal}}
\end{figure}

Before embarking on a MC study of this problem it is useful to
establish the results of a simple mean field study of B-B' order
(at equilibrium) for varying $x$. Fig.8 shows the result of such
a calculation, done via the mapping of the lattice gas to an
Ising model at constant `magnetisation' (which corresponds
to the difference in concentration of B and B'). 
One can capture this
effect within mean field theory by considering an AF 
Ising model
at different magnetic fields,
and computing the AF order parameter, {\it i.e}, the ordering peak,
and the magnetisation.  
As seen in Fig.8, the $\{ \pi, \pi\}$ ordering
at low temperatures decreases
uniformly with increasing magnetization. However, it truly
vanishes only when the magnetization
is unity, correponding to a purely B (eg. Fe) or
B' (eg. Mo) compound.
This does not correspond to the experimental situation.

\begin{figure}[h]
\centerline{
\includegraphics[width=6.0cm,height=4.5cm,clip=true]{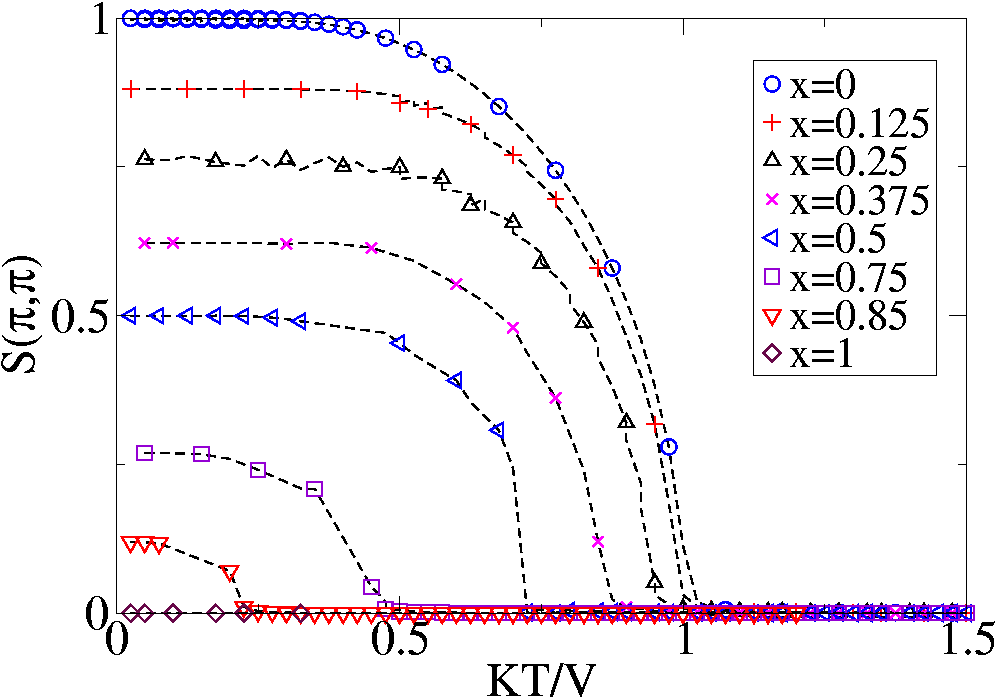}}
\caption{Structure factor at ($\pi,\pi$) from mean field theory for
 different B-B' proportions}
\end{figure}

Next we did  Monte Carlo
simulations for the Ising model for different non-zero
values of magnetization. The results
for cooling runs are shown in Fig.9.
The $\{ \pi, \pi\}$ ordering is found to almost vanish
at $x=0.5$, much before the mean
field  prediction. Interestingly, the experimental
XRD data at $x=0.5$ does not seem to
show the $19.6^{\circ}$ peak related to ordering either.

\begin{figure}[h]
\centerline{
\includegraphics[width=6.0cm,height=4.5cm,clip=true]{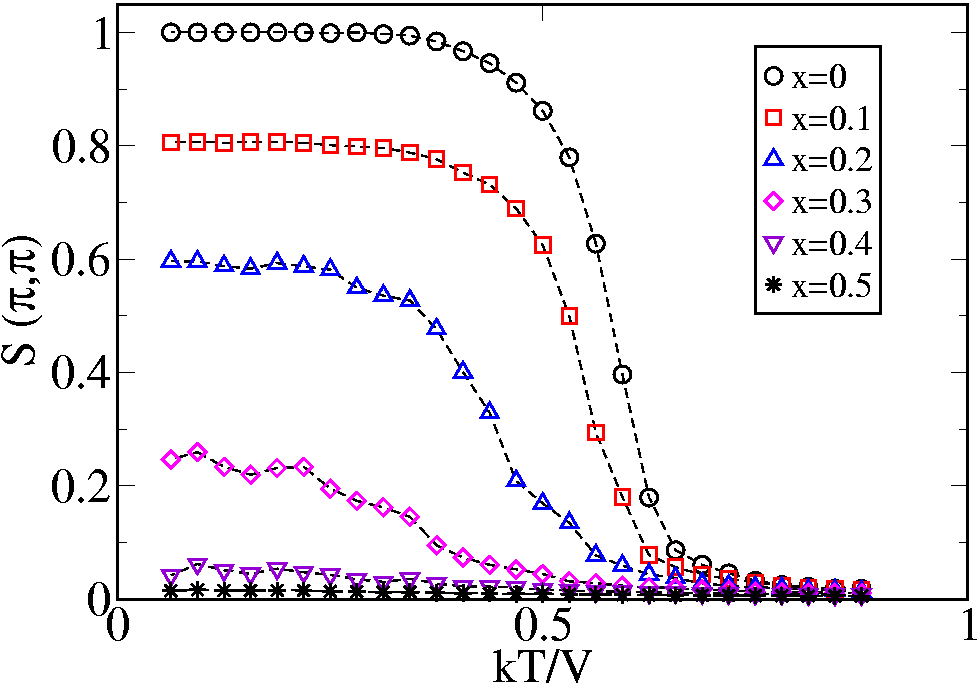}}
\caption{
Structure factor vs T (cooling runs) for different
B-B' proportions.
}
\end{figure}

For any specific proportion, $x=0.1$, say, the variation 
of $S(\pi, \pi)$ with $T_{ann}$
is shown in Fig.10, for various $\tau_{ann}$.
The overall behaviour is similar to that for equal
B-B' proportions, namely that the non-monotonicity
decreases upon increasing annealing time, and gradually
approaches the equilibrium
cooling curve. Since the equilibrium saturation
value of the $(\pi,\pi)$ structure factor
in this case is less than 1, hence the maximum values of
the non-equlibrium curves for any
given value of MC steps is less than the corresponding value
for the equal proportion case.

In Fig.11, on the other hand,
the annealing time is kept fixed,
while the proportions are varied.
While $S(\pi, \pi)$ obviously has  a nonmonotonic
dependence on $T_{ann}$, the maximum values progressively increase
with decreasing  $x$ as expected. In the
case of annealing with memory, as shown in
Fig.12,
the behaviour is once again nonmonotonic, except that the
maximum of the structure factor
nearly always hits the equilibrium value.

\begin{figure}[h]
\centerline{
\includegraphics[width=6.0cm,height=4.5cm,clip=true]{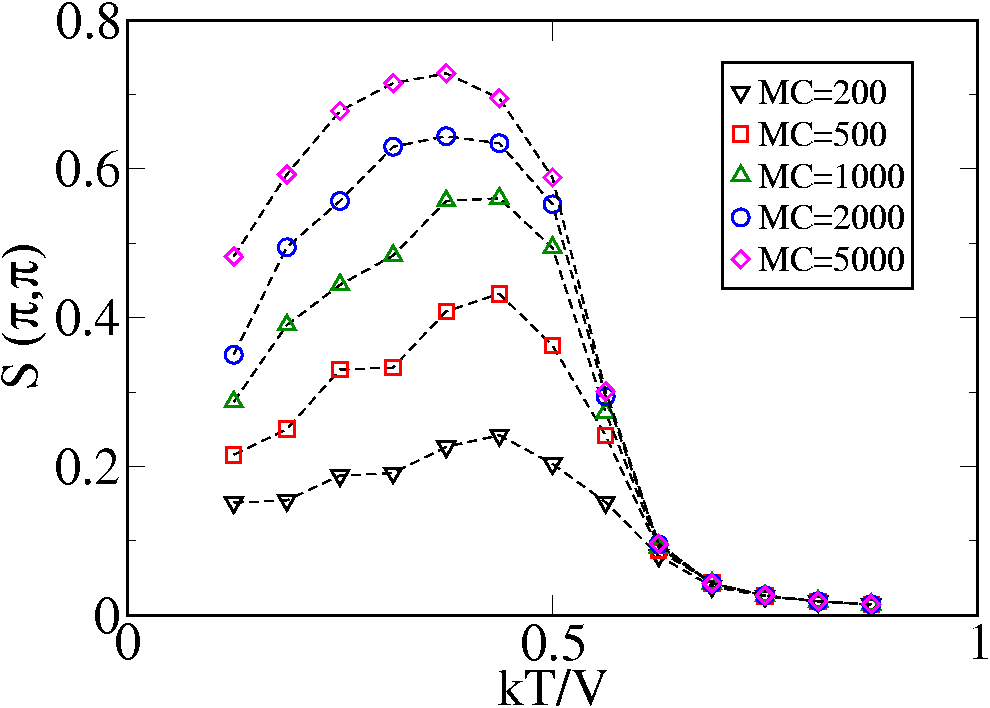}}
\caption{
Structure factor vs T (heating run without memory) for  $x=0.1$,
 different Monte Carlo steps/spin. Average over 300 initial configurations.
}
\end{figure}

\begin{figure}[h]
\centerline{
\includegraphics[width=6.0cm,height=4.5cm,clip=true]{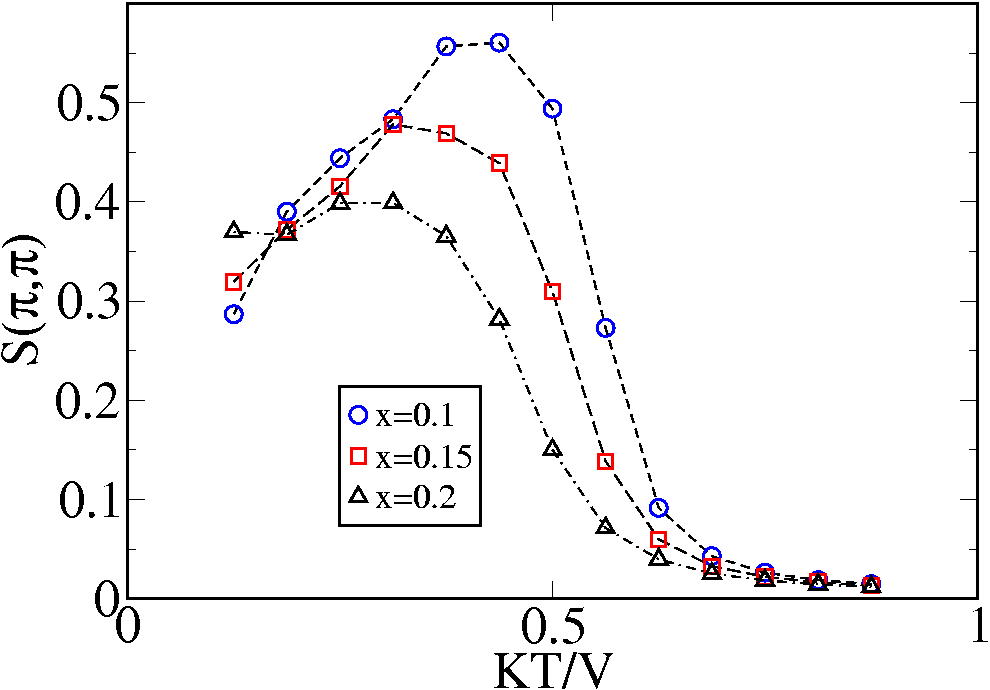}}
\caption{
Structure factor vs T (heating run without memory) for different $x$,
 fixed Monte Carlo steps/spin=1000.
}
\end{figure}

\begin{figure}[h]
\centerline{
\includegraphics[width=6.0cm,height=4.5cm,clip=true]{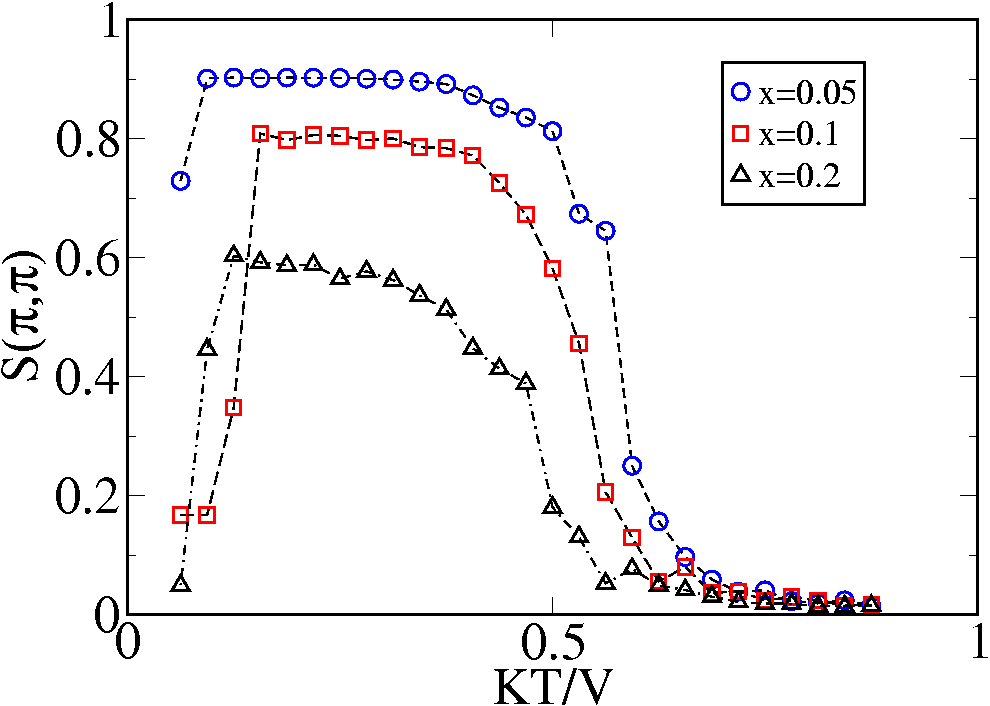}}
\caption{
Structure factor vs T (heating run with memory) for different $x$,
 fixed Monte Carlo steps/spin=1000.
}
\end{figure}

Fig.13, left,  shows a low temperature 
configuration in a well
equilibriated sample  
for 2:1 proportion of B:B'  
species. The ordering is 
disturbed by the occurence of large antisite patches, 
forced by  the excess B. However, the species which occurs in the lesser 
proportion is found to order
as much as possible, and does not form nearest neighbours. 
In this `phase-colouring' scheme, the antisite regions show up as checkerboard patterns.
The actual Fe-Mo checkerboad regions, on the other hand, shows up as domains of a particular colour. The black and white domains are out of phase with respect to each other, while they
are separated by thick 'antisite walls'.
 
Fig.13, right, shows the configuration 
for B:B'=3:1, the concentration at which
the ordering all but disappears. It is observed that the 
antisite patches have started 
percolating at this concentration.

\begin{figure}[h]
\centerline{
\includegraphics[width=3.5cm,height=3.5cm,angle=270,clip=true]{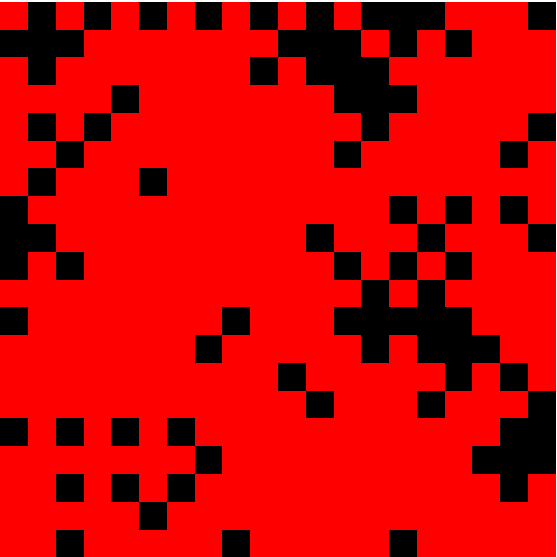}
\hspace{.2cm}
\includegraphics[width=3.5cm,height=3.5cm,angle=270,clip=true]{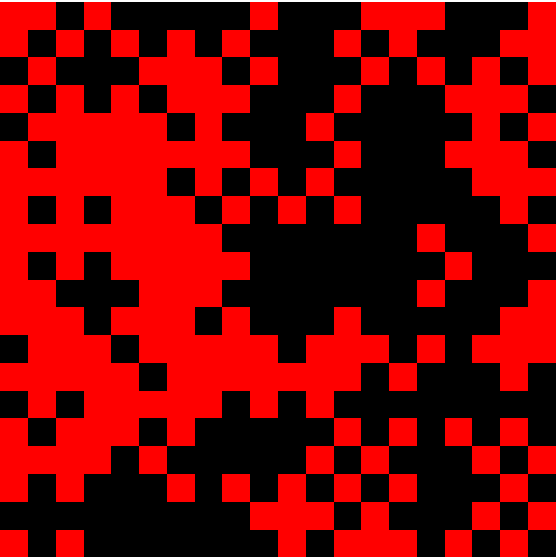}}
\caption{Left: Equilibrium low T (=0.5) configuration for large
MCsteps=5000, Fe:Mo=2:1 (x=0.33). Right:
Equilibrium low T (=0.5) configuration for large MC
steps=5000, Fe:Mo=3:1 (x=0.5).}
\end{figure}

\section{Computational checks}

While the previous sections essentially summarize the results for 
the Fe-Mo ordering
problem, it is necessary to probe the sensitivity of the results to 
the system size
and the updating cluster size chosen. In this section, we study these 
systematics, thereby
providing computational indicators to the robustness of our results. 
In Fig 14 and 15, we plot
the dependence of domain sizes on system size and B-B'  ratio. The three 
columns
stand for L=20,40 and 80 from left to right, while the three rows 
correspond to B-B'
ratio 1:1,1:2 and 1:3 from top to bottom. It is found for all the 
proportions that the rough
domain size remains the same irrespective of the system size, which 
means that there are more
domains for larger system sizes, with more structure to them. 
This is precisely the reason
for the sharp suppression of the nonequilibrium structure factor 
with increasing system size
obseved in Section V. 

\begin{figure}[t]
\centerline{
\includegraphics[width=7.5cm,height=7.5cm,clip=true]{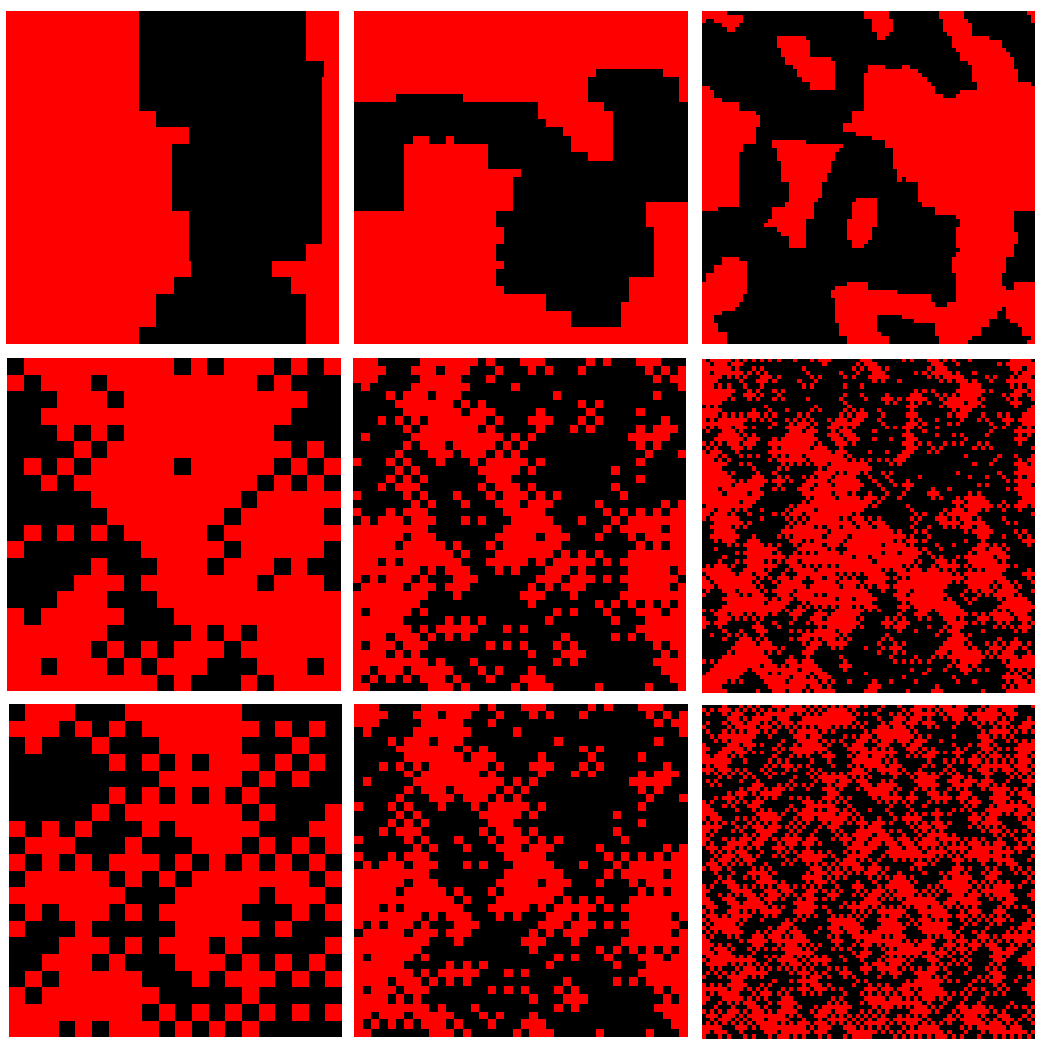}}
\vspace{.2cm}
\caption{
Domain structure for various system sizes: $L=20,~40,~80$ (left to
right) and B:B ratio $1:1,~1:2,~1:3$ (top to bottom). This
is a poorly equilibriated system, with $\tau_{ann} =200$MCS.
Notice the increase in antisite disorder with increasing
B:B' disproportionation, and the rough constancy of the
typical domain size with increasing $L$.
}
\end{figure}

\begin{figure}[t]
\centerline{
\includegraphics[width=7.5cm,height=7.5cm,clip=true]{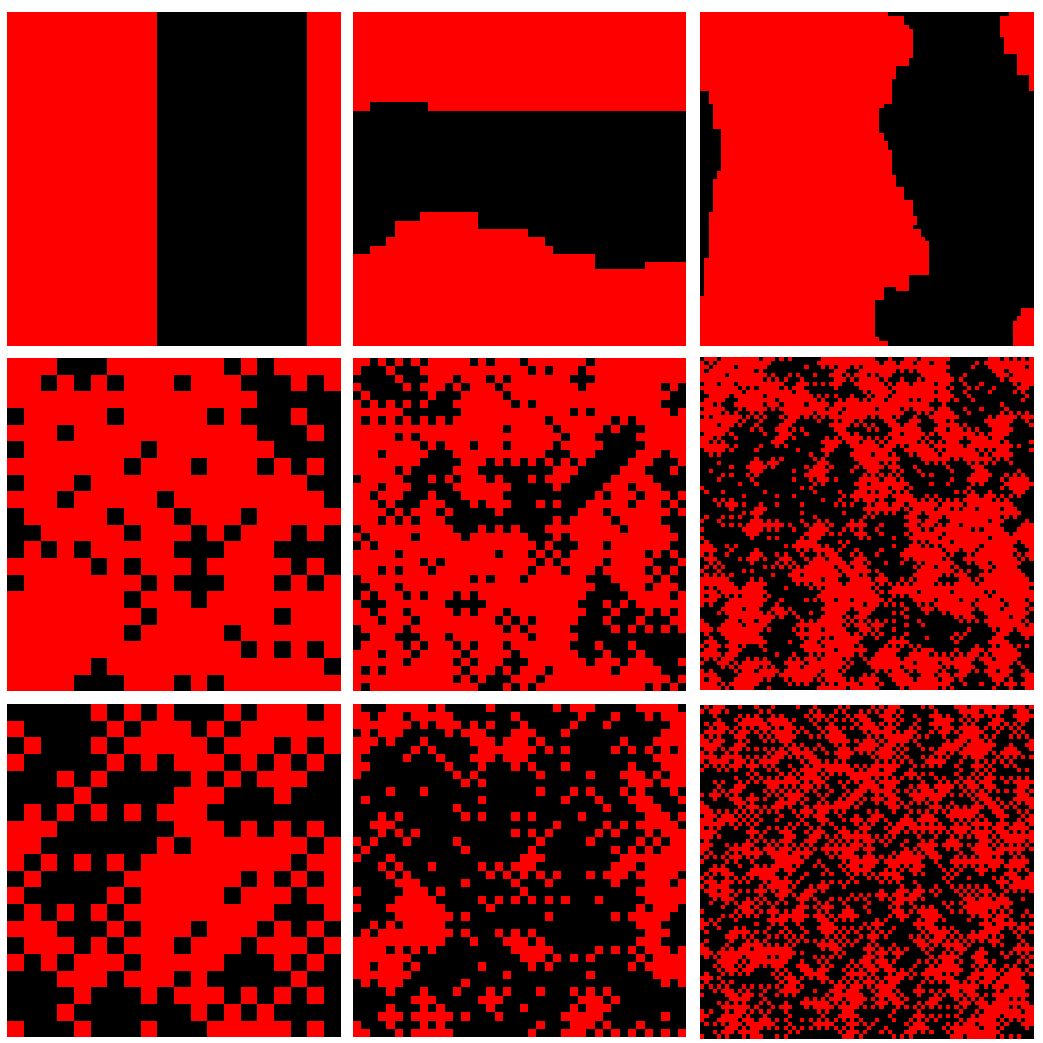}}
\vspace{.2cm}
\caption{
The same as Fig.14, but with much longer equilibriation,
$\tau_{ann} =5000$MCS.
}
\end{figure}

The difference between Fig.14 and Fig.15 lies in the fact that the 
former shows configurations 
obtained after 200 Monte Carlo steps, while the latter shows 
configurations obtained by continuing the same run upto 5000 MCS. 
It is observed that for the case of equal proportions, the
domains tend to clump together and become more uniform 
as the number of MCsteps increase.
For the case of different proportions especially Fe:Mo=2:1, i
the fraction of antisite defects for lesser MCS is substantially 
larger than that for greater MCS. This is due to the fact that
 for MCS=200, there is an extra source of antisite defect 
formation in addition to the
 disproportionation, namely insufficient equilibriation. 
In other words, even the 
 species which occurs in lesser proportion forms nearest 
neighbours, in contrast to
 the equilibrium cases of Fig.13. On the other hand, 
for Fe:Mo=3:1,
 the inherent disorder due to disproportionation is so 
large as to mask out the additional
 nonequilibrium effects. 

 In Fig.16, the domain pattern is shown for two different 
update  cluster sizes, $L_{c}=4$
 (first column)  and $L_{c}=10$ (second column). 
The first row corresponds to 200 MCS, while
 the second corresponds to 5000 MCS. No appreciable 
qualitative change is observed, indicating
 the robustness with respect to this parameter, 
at least within the regimes considered.

\begin{figure}[h]
\centerline{
\includegraphics[width=5.0cm,height=5.0cm,clip=true]{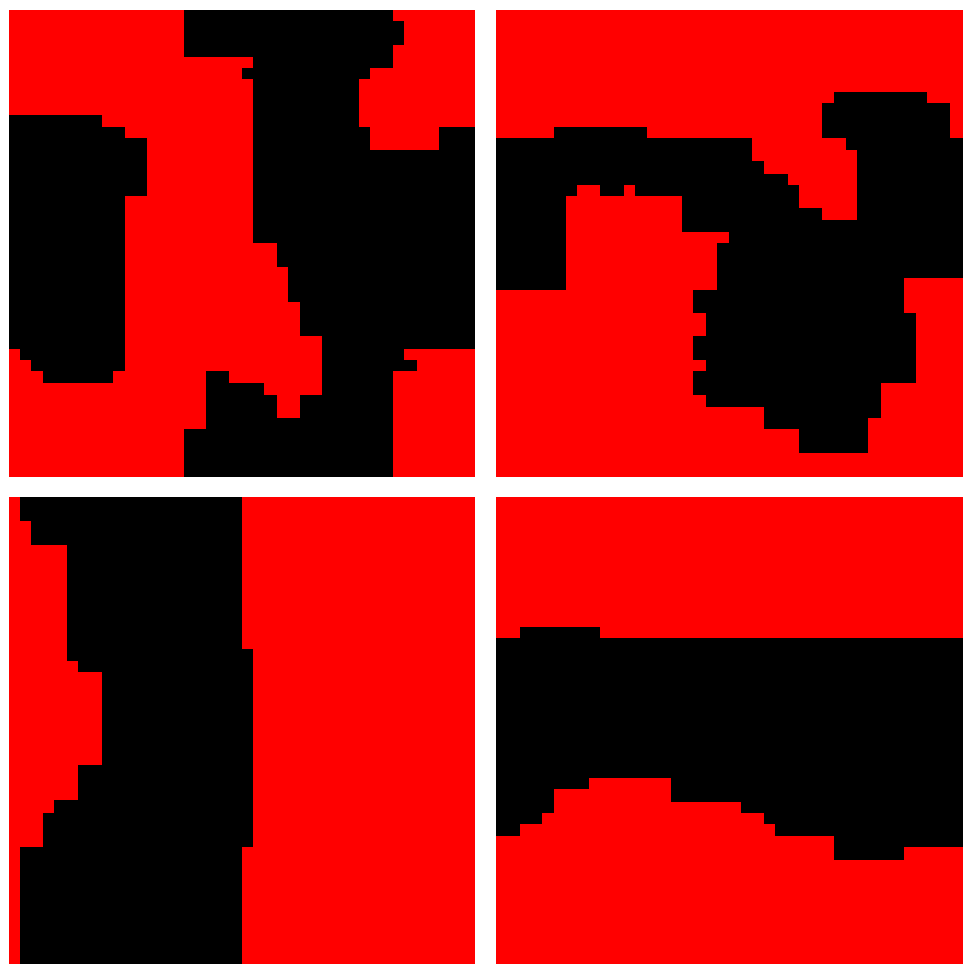}}
\vspace{.2cm}
\caption{
Domain pattern with two different update cluster sizes: 1st column: $L_{C}=4$,
2nd column: $L_{C}=10$. 1st row: 200 MCS, 2nd row: 5000 MCS. 
$L=40$. The same initial configuration were used as in
Fig.14 and Fig.15.
}
\end{figure}

\section{Ordering in ternary B-B'-B'' systems}

\begin{figure}[h]
\centerline{
\includegraphics[width=6.0cm,height=5.0cm,clip=true]{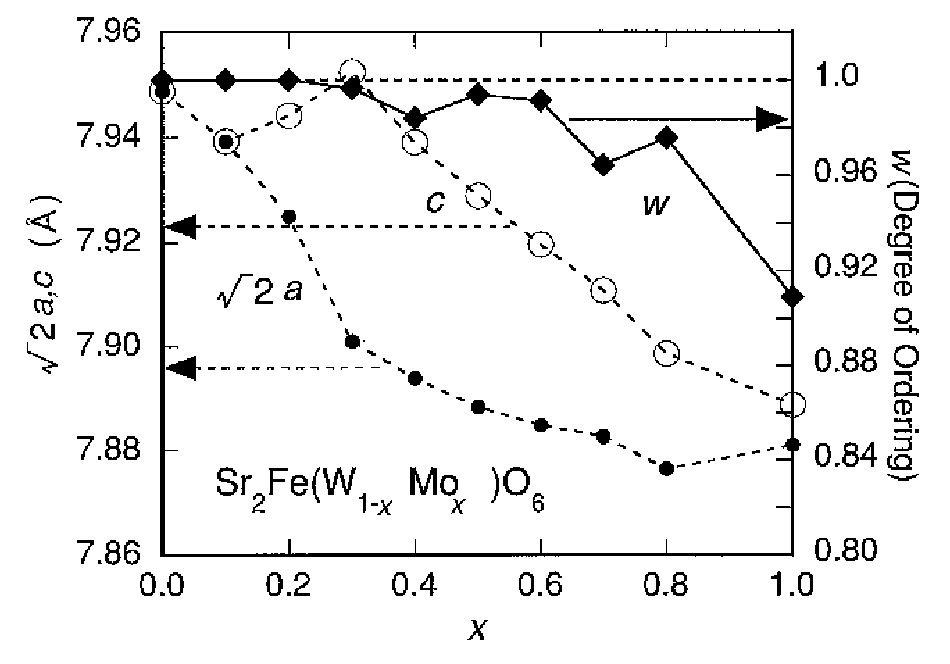}}
\vspace{.2cm}
\caption{
Experimental data on B-B$^{\prime}$ site ordering in 
Sr$_{2}$FeMo$_{x}$W$_{1-x}$O$_{6}$;
figure taken from Ref~\cite{TokuraFeMoW}
}
\label{Tokura_expt}
\vspace{.2cm}
\end{figure}

\begin{figure}[h]
\centerline{
\includegraphics[width=6.0cm,height=4.5cm,clip=true]{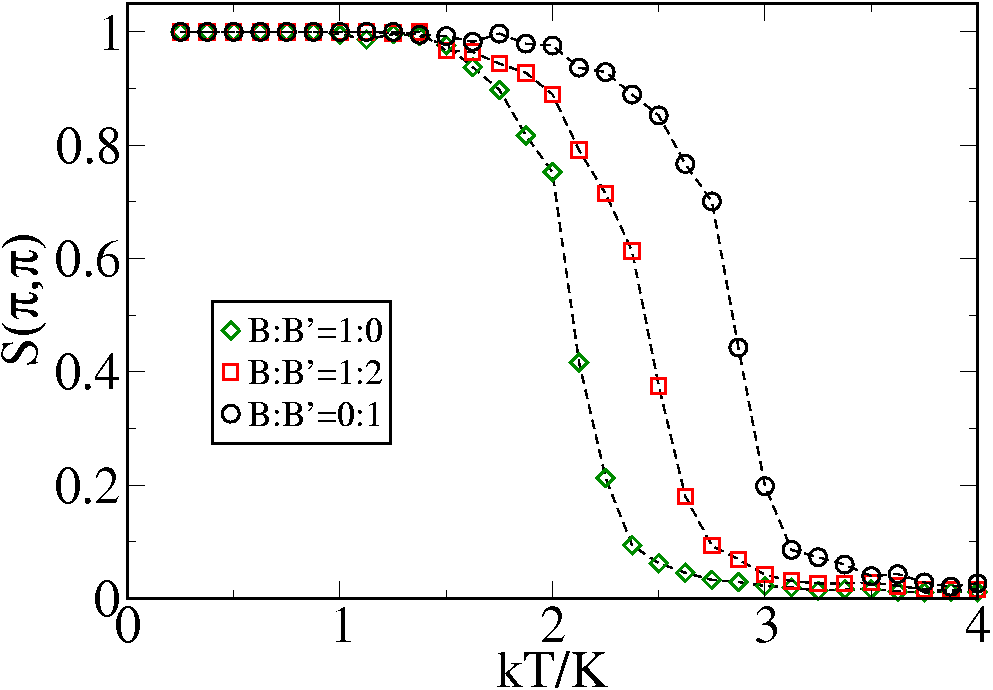}}
\vspace{.2cm}
\caption{
Structure factor vs T (cooling runs) for different Mo-W
proportions for the 3species case.
}
\label{3species_cooling}
\end{figure}

\begin{figure}[h]
\centerline{
\includegraphics[width=6.0cm,height=4.5cm,clip=true]{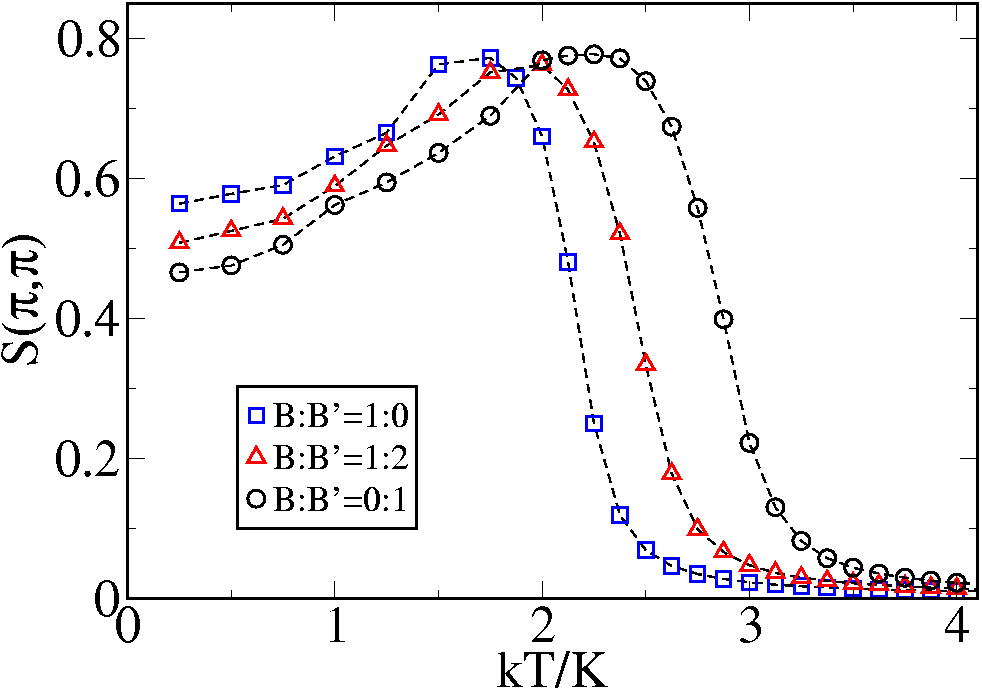}}
\vspace{.2cm}
\caption{
Structure factor vs T (heating without memory) for different
Mo-W proportions for a fixed annealing time (2000 MCsteps/spin).
}
\label{3species_heating_fixed_MCS}
\end{figure}

\begin{figure}[h]
\centerline{
\includegraphics[width=6.0cm,height=4.5cm,clip=true]{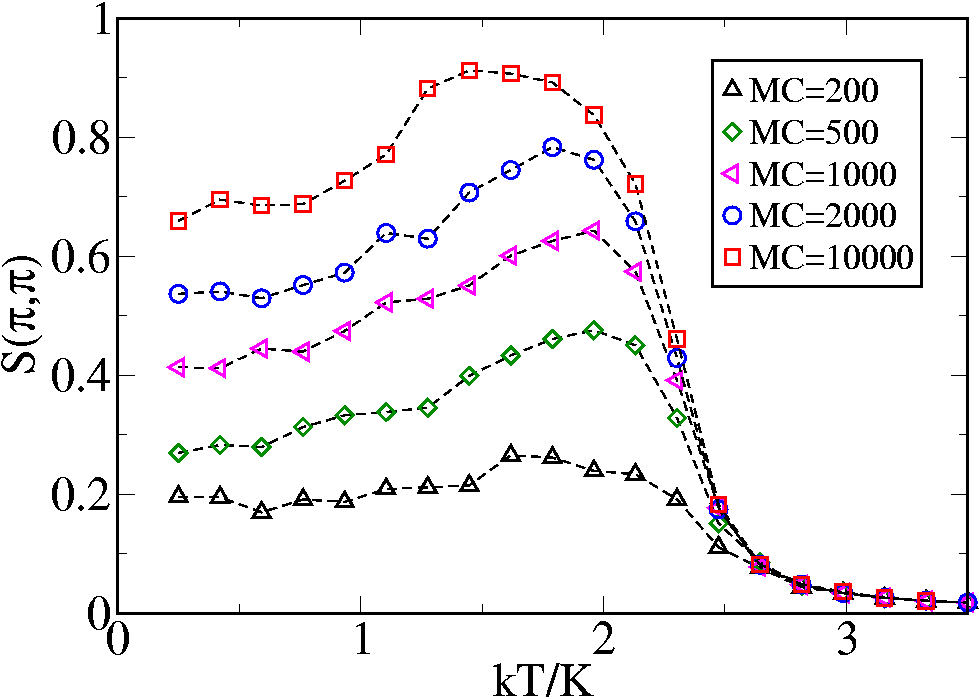}}
\vspace{.2cm}
\caption{
Structure factor vs T (heating without memory)
for Mo:W=1:1 for different annealing times.
}
\label{}
\label{3species_heating_fixedMoW}
\end{figure}

\begin{figure}[h]
\centerline{
\includegraphics[width=5.0cm,height=4.5cm,clip=true]{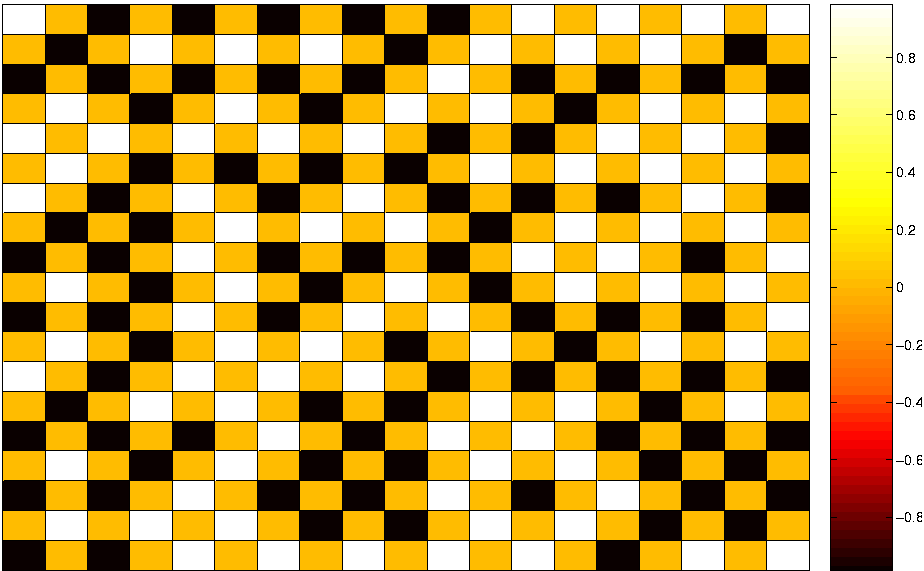}}
\vspace{.2cm}
\caption{
Equilibrium low T configuration for Mo:W=1:1. The black
squares represent the Mo, while the white and the yellow represent
the W and Fe respectively.
Notice perfect B-B'  ordering,
 but no ordering on the $B^{\prime}$ site between Mo and W.
}
\label{3species_config}
\end{figure}

If Sr$_{2}$FeMoO$_{6}$ is doped with tungsten to create the 
series of compounds 
Sr$_{2}$FeMo$_{x}$W$_{1-x}$O$_{6}$, then one has a 
compositional/structural problem
involving three species: Fe, W and Mo. In experiments done by 
Kobayashi {\it et.al.}~\cite{TokuraFeMoW} it was found that the 
lattice parameters $a$
and $c$ increase with increasing W concentration, indicating 
a size mismatch. However,
the more interesting observation appears to be on the ordering of
Fe with respect to Mo/W.
Both Kobayashi {\it et. al.} and Sarma {\it et.al.}~\cite{SugataFeMoW} 
observed a larger
degree of B-B'
ordering in the heavily tungsten doped compounds. The ordering 
is indeed observed by Kobayashi
{\it et. al.} to be almost $100\%$ for $x\leq0.6$, as seen in Fig~\ref{Tokura_expt}.
 However, no 
such ordering is observed on the
 B'  site for the W and Mo species. Instead, a random mixing 
of W and Mo is maintained throughout the series.

The above results suggest that there is an effective short range 
attractive interaction
between Fe and either W or Mo, while there is only a
much weaker interaction  
between W and Mo.
However, the Fe-W attractive tendency is apparently larger than 
the Fe-Mo one, since the
degree of ordering increases with increasing W concentration. 
Hence, we model this three-species
annealing problem in the following manner. 

We consider a site-variable $S$ which can take three 
possible values: 0 (Fe), 1 (W) and 
-1 (Mo). Then, we consider the following model hamiltonian:
\begin{equation}
H_{eff}=\sum_{<ij>}[JS_{i}S_{j}+KS_{i}^{2}S_{j}^{2}+L(S_{i}^{2}S_{j}+S_{j}^{2}S_{i})]
\end{equation}
This model, a variant of the Blume-Emery-Griffiths 
model~\cite{Blume}, 
 is a generalization of the Ising model to spin $S=1$ rather
than 1/2. One can reduce this model 
to an effective Ising model in the binary
 limits of (Fe,Mo), (Fe,W) and (Mo,W), with effective 
exchange couplings given by $J+K-2L$,
$J+K+2L$ and $J$ respectively. Motivated by the experimental 
observations, we choose the
 values of the parameters $J,K$ and $L$ such that the 
effective energy scale for Fe-W
 ordering is somewhat larger than the Fe-Mo scale, both being 
substantially large compared to
 the Mo-W ordering scale. With such a parameter choice, 
a Monte Carlo is performed as before,
 and the extent of B-B'  site ordering 
(not distinguishing between Mo and W on the
 B' site, is again quantified by the structure 
factor at $\{ \pi,\pi \}$. The variation
 of this structure factor with temperature for different 
relative Mo-W concentrations is plotted
 in Fig.~\ref{3species_cooling} for cooling runs. 
The extreme W-only case of course exhibits
 the highest $T_{c}$ while the Mo-only case shows the 
lowest one, as expected. For 
 intermediate Mo-W concentrations, the $T_{c}$ for the 
3-species assembly interpolates between
 the two limits. Hence, if one assumes that 
one has annealed the sample long 
 enough to reach equilibriation at any specific annealing 
temperature, the B-B' 
 ordering increases uniformly with increasing W concentration, 
as observed in the experiment.
 However, it is interesting to note that the behaviour 
gets reversed in the off-equilibrium 
 non-monotonic case, as observed in our result for heating 
run without memory given in
 Fig.~\ref{3species_heating_fixed_MCS}. This is because, at 
any given low temperature,
 the activation probability for the lowest $T_{c}$ 
 compound 
is the highest, while that with the highest $T_{c}$ is the 
lowest. Hence, the compound with the lowest $T_{c}$ reaches 
equilibrium fastest. This is the reason for the reversal. 
For completenes, the 
  heating run without memory data is also shown in 
Fig.~\ref{3species_heating_fixedMoW} for different annealing times, keeping the Mo-W 
proportion fixed at 1:1. The same
 transition from non-monotonic to montonic behaviour 
is observed in this case also.
 Finally, Fig.~\ref{3species_config} shows a ground state 
equilibrium configuration for Mo:W=1:1
 obtained from a cooling run. It is observed that there 
is perfect ordering on the
 B site, corresponding to the Fe sublattice. 
On the B'  site, Mo and W compete
 for space. There appears to be no significant ordering 
amongst these latter. Hence, overall
 this simulates the experimental situation 
for Sr$_{2}$FeMo$_{1-x}$W$_{x}$O$_{6}$ 
 quite effectively.

\section{Conclusion}

We have shown that antisite disorder can occur due to 
three reasons: (i)~insufficient annealing
at low temperatures leading to non-equilibrium configurations, 
(ii)~sufficient annealing leading
to equilibrium configurations but at high temperatures 
close to the order-disorder transition,
and (iii)~sufficient annealing leading to equilibrium 
configurations at low temperature, but with
different concentrations for the B and 
the B' species.

\vspace{.2cm}

{\it Acknowledgements:} We thank D. D. Sarma for introducing us to this
problem and several discussions. We also thank Sugato Ray, Anamitra Mukherjee,
and Sanjeev Kumar for discussions. We acknowledge use of the Beowulf Cluster
at HRI. 

\vspace{.4cm}

\section{Appendix}
We provide the mapping between the lattice gas model and the
Blume-Emery-Griffiths model here.
If we represent a three-component lattice gas of species A,B and 
C using spin S=1, with components 1,0,-1, then the number of 
atoms of each type is given by:
\begin{eqnarray}
N_{A}&=&\frac{1}{2}\sum_{i=1}^{N}(S_{i}^{2}+S_{i})  \cr
N_{B}&=&\frac{1}{2}\sum_{i=1}^{N}(S_{i}^{2}-S_{i}) \cr
N_{C}&=&\sum_{i=1}^{N}(1-S_{i}^{2}) 
\end{eqnarray}

An effective spin model can be written as:

\begin{equation}
H=\sum_{ij}[JS_{i}S_{j}+KS_{i}^{2}S_{j}^{2}+L(S_{i}^{2}S_{j}+S_{j}^{2}S_{i})]
\end{equation}
where the three parameters $J,K,L$ above are related to the 6 
interspecies nearest neighbour
interactions  by: 
\begin{eqnarray}
J&=&E_{AA}/4+E_{BB}/4-E_{AB}/2 \cr
K&=&E_{AA}/4+E_{BB}/4+E_{CC}+E_{AB}/2-E_{AC}-E_{BC} \cr 
L&=&E_{AA}/4-E_{BB}/4-E_{AC}/2+E_{BC}/2 
\end{eqnarray}

It is to be noticed that the relation between the `Ising' parameter 
$J$ with the energies 
involve only the A and B species, and is identical to the 
corresponding relation for the
binary alloy case. The explicit relation between the Ising parameter 
J and the lattice gas 
parameter $V$ is given by $V=4J$ provided the bonds are counted only once 
(i.e., the summation over
$ i,j$ is restricted to $i<j$). Hence, in terms of a lattice gas model, 
the three parameters 
 $E_{AA}, E_{BB}$ and $E_{AB}$ can be expressed in terms of a single parameter:
$$V=(E_{AA}+E_{BB}-2E_{AB})$$.


\begin{thebibliography}{9}
\bibitem{dp-rev} D.D. Sarma, Current Op. Solid St. Mat. Sci.,{\bf 5}, 261 (2001).
\bibitem{KimCheong} T.H. Kim, M. Uehara, S. W. Cheong and S. Lee, Appl. Phys. Lett.,{\bf 74}, 1737 (1999).
\bibitem{Martinez} B. Martinez, J. Navarro, Ll. Balcells and J. Fontcuberta, J. Phys.: Condens. Matter {\bf 12},
 10515 (2000).
\bibitem{GarciaLanda} B. Garcia Landa {\it et al}, Solid State Comm., {\bf 110}, 435 (1999).

\bibitem{Kobayashi} K.-I. Kobayashi, T. Kimura, H. Sawada, K. Terakura and Y. Tokura, Nature {\bf 395},
 677 (1998).
\bibitem{staufferbook} D. Chaudhuri and D. Stauffer, {\em Principles of
Equilibrium Statistical Mechanics}  
\bibitem{ord-expt-prl} D.D. Sarma, S. Ray, K. Tanaka, M. Kobayashi, A. Fujimori, 
P. Sanyal,
 H.R. Krishnamurthy and C. Dasgupta, Phys. Rev. Lett.,{\bf 98}, 157205 (2007).
\bibitem{topwal} D. Topwal, D.D. Sarma, H. Kato, Y. Tokura and M. Avignon, Phys. Rev. B.,
{\bf 73}, 094419 (2006)

\bibitem{TokuraFeMoW} K.-I. Kobayashi, T. Okuda, Y. Tomioka,T. Kimura and Y.Tokura,
J. Magn. Magn. Mat., {\bf 218}, 17 (2000).
\bibitem{SugataFeMoW} S. Ray, A. Kumar, S. Majumdar, E.V. Sampathkumaran, D.D. Sarma,
J. Phys.: Condens. Matter {\bf 13}, 607 (2001). 
\bibitem{Blume} Mukamel and Blume, Phys. Rev. A, {\bf 10}, 610 (1974).

\end{thebibliography}
\end{document}